\documentclass[dvips]{article}

\usepackage{icrc2011}

\title{Detection of TeV emission from the intriguing composite SNR G327.1-1.1 }

\newcommand {\snr} {G327.1$-$1.1}
\newcommand {\h} {H.E.S.S.}
\newcommand {\g} {$\gamma$}
\newcommand {\xmm} {XMM-\textit{Newton}}

\newcommand{\etal}{\MakeLowercase{\textit{et al. }}} 
\shorttitle{Acero \etal TeV detection from the composite SNR  G327.1-1.1 }

\authors{F. Acero$^{1}$, A. Djannati Ata\"i $^{2}$, A. F\"orster$^{3}$, Y. Gallant$^{1}$, M. Renaud$^{1}$  for the H.E.S.S. collaboration}
\afiliations{$^1$LUPM, CNRS/IN2P3, Universit\'e Montpellier II, Place Eug\`ene
Bataillon, 34095 Montpellier Cedex 5, France\\
 $^2$Astroparticule et Cosmologie, CNRS/Universit\'e Paris 7, Batiment Condorcet, Paris 75205 cedex 13, France\\
$^3$Max-Planck-Institut f\"ur Kernphysik, Saupfercheckweg 1, 69117 Heidelberg, Germany }
\email{facero@in2p3.fr } 

\abstract{The shock wave of supernova remnants (SNRs) and the wind termination shock in pulsar wind nebula (PWNe) are 
considered as prime candidates to accelerate the bulk of Galactic cosmic ray (CR) ions and electrons. The SNRs hosting a PWN (known as composite SNRs) provide excellent laboratories to test these hypotheses.
The SNR G327.1--1.1 belongs to this category and exhibits a shell and a bright central PWN, both seen in radio and X-rays. Interestingly, the radio observations of the PWN show an extended blob of emission and a curious narrow finger structure pointing towards the offset compact X-ray source  indicating a possible fast moving pulsar in the SNR and/or an asymmetric passage of the reverse shock.
We report here on the observations, for a total of 45 hours, of the SNR G327.1--1.1 with the H.E.S.S. telescope array which resulted in the detection of TeV \g-ray emission
 in spatial coincidence with the PWN. }

\begin{document}
\maketitle

\section{Introduction}

The survey of the Galactic plane conducted by the  \h~system of telescopes since 2004 has led to the discovery of TeV gamma-ray emission from more than 60 sources in the Galaxy \cite{gast11}. Although the nature of a large part of theses sources remains still unknown, the identification of a large sample of pulsar wind nebulae (PWNe) and supernova remnants (SNRs, either shell-type or in interaction with molecular clouds) has provided significant progress in constraining and understanding the acceleration mechanisms at play, and the physical conditions within cosmic sources (see e.g. \cite{hinton09}). 
The determination of the origin of the TeV emission can be particularly difficult for composite SNRs (i.e. those hosting a PWN) when their angular size is smaller or close to the point spread function (PSF) of the instrument  : the TeV emission could stem either from the shell of the remnant and/or from the PWN.
We report here on the discovery of TeV $\gamma$-rays from the composite SNR \snr~using the \h~telescopes. 

\h~is an array of four imaging atmospheric Cherenkov telescopes located in the Khomas Highland in Namibia at an altitude of 1800 m above sea level.  Each telescope has a mirror area of 107 m$^2$ \cite{bernlohr03}, a the total field of view of $5^\circ$  equipped with a camera consisting of 960 photomultiplier tubes \cite{vincent03}.  The stereoscopic detection and reconstruction of events allows a good angular precision per event $(0.1^\circ)$,  a good energy resolution ($15 \%$ on average),  and a high background rejection efficiency.  Detailed information on H.E.S.S. is given in \cite{hinton04} and the most recent
developments in Cherenkov reconstruction and rejection methods  are described in  \cite{ohm09,dr09,becherini11}.

\section{The case of the SNR \snr}

The SNR \snr~belongs to the category of composite SNRs as it exhibits in the MOST SNR radio survey at 843 MHz \cite{whiteoak96} both a faint shell
of relatively small angular size ($r=0.15^{\circ}$) and a bright PWN, slightly off-center with respect to the SNR shell.  
An intriguing narrow finger-like radio feature is also seen to the North-West of the brightest region of the PWN,  pointing towards a compact X-ray 
source considered as the putative pulsar powering the PWN.

Recent high spatial resolution \xmm~and \textit{Chandra} X-ray observations of  \snr~have revealed that the compact source is embedded in a
cometary structure, exhibiting non-thermal emission,  aligned with the radio finger feature \cite{temim09}.  
Such morphology could be due to a high velocity of the pulsar \cite{swaluw04} and/or to the inhomogeneities of the ambient medium in which the SNR
is expanding and which can lead  to an asymmetric passage of the reverse shock crushing the PWN \cite{blondin01}.
Concerning the properties of the pulsar candidate, no pulsations have been detected so far, neither in radio,  in X-rays \cite{temim09} nor in
 GeV \g-rays in the 2nd \textit{Fermi}-LAT catalog \cite{fermi2cat}.
 However, based on empirical relationships between the nebula's non-thermal X-ray luminosity and the current spin-down energy loss rate, \cite{temim09} have estimated  the latter, $\dot{E} \sim 3 \times 10^{37}$ erg s$^{-1}$, and the pulsar period, $P \sim 35$ ms,
assuming a SNR age of 18 kyrs and a distance of 9 kpc (as derived by \cite{sun99}).

The X-ray observations have also revealed diffuse thermal X-ray emission from the shell of the SNR which coincides  
well with the radio contours \cite{temim09}.

\section{H.E.S.S observations and analysis}
\label{sec:gamma}

The region of the SNR \snr~has been observed by \h~between
2005 and 2008 during the Galactic plane scan for 15h and then for 30h of dedicated observations in 2009 and 2010
with zenith angles ranging from  30$^{\circ}$ to 43$^{\circ}$ with a mean value of  33$^{\circ}$.
The resulting total \h~observation time for this target is of 45 hours after data quality cuts.

The data set was analyzed using the \textit{Model analysis} \cite{dr09} which exploits the
full pixel information by comparing the recorded shower images with a pre-calculated shower model
using log-likelihood minimization. 
For the image generation an additional cut on the event's reconstructed direction uncertainty was applied ($dDir\leq0.03^{\circ}$).
The effects of this additional cut on the angular resolution and the effective area are discussed in Sect. 4.9 of \cite{dr09}.
This resulted in an improved average angular resolution for this data-set  ($r_{\rm 68\%} = 0.05^{\circ}$) 
allowing us to precisely locate the source position at the expense of a smaller statistics.
For the spectral analysis, the standard cuts were used in order to use the full statistics available.

 All results presented were cross-checked with a multivariate analysis \cite{ohm09} using an independent calibration and gamma/hadron separation,
  which yielded consistent results. 
  
\begin{figure}
   \centering

   \includegraphics[bb= 65 220 465 575, clip,width=7.cm]{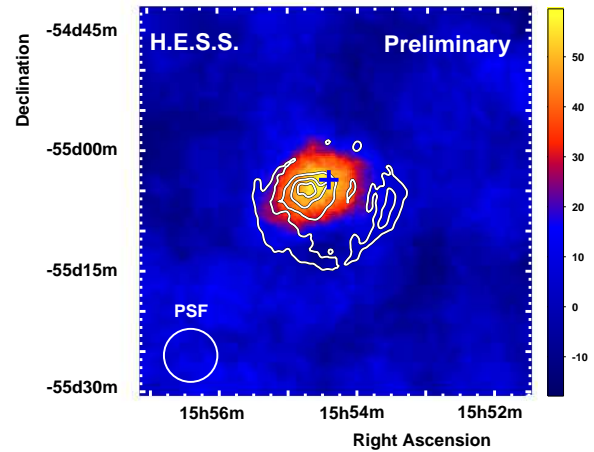}   
   \includegraphics[bb= 473 145 515 581, clip,width=0.59cm]{icrc0928_fig01.ps}
   
      \vspace{-0.4cm}

   \caption{TeV \g-ray excess map ($0.8^{\circ} \times0.8^{\circ}$) of the region of the SNR \snr~produced using an oversampling radius  of 0.08$^{\circ}$ 
   together with the radio contours from the  843 MHz MOST survey overlaid in white.
    The average \h~68\% containment radius for the dataset is shown by the white circle and
    the position of the compact X-ray source (the putative pulsar) with a blue cross. 
    The transition between blue and red in the color scale is at the level of 4$\sigma$.}
   \label{fig:excess}
\end{figure}

\subsection{Morphological results}

Preliminary H.E.S.S. results of these observations unveiled a compact region of TeV \g-ray emission detected
at 8$\sigma$. The corresponding  \h~excess map, shown in Fig.~\ref{fig:excess},  reveals that the 
\g-ray emission is contained in the interior of the remnant, where the bright PWN is observed both in the 
radio and the X-ray band. It is also interesting to note that no \g-ray emission is detected from the shell 
of the remnant.

\begin{figure}
   \centering

   \includegraphics[width=7.cm]{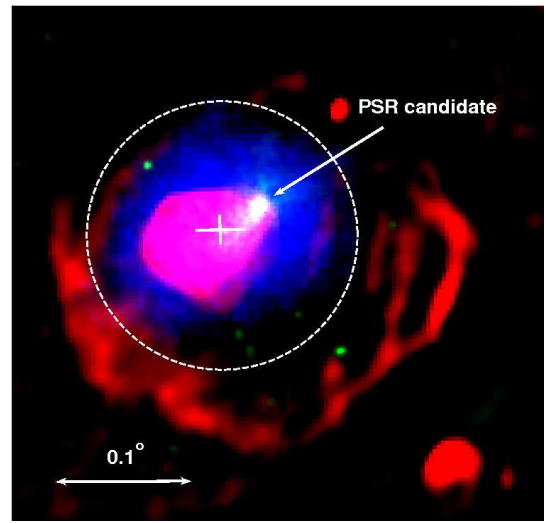}   
 
      \vspace{-0.2cm}

   \caption{ Multi-wavelength view of the SNR \snr. The radio (MOST survey), X-ray (\xmm~in the 0.5-6 keV energy band) and
   \g-ray map are respectively represented in red, green and blue color scales. The white cross represents the best-fit centroid of 
   the \g-ray emission and the associated errors for each coordinate. The white circle illustrates the circle of radius 0.1$^{\circ}$ used as spectral extraction region. }
   \label{fig:rgb}
\end{figure}

In order to derive the centroid position and intrinsic size of the \g-ray source, a two-dimensional Gaussian fit was performed
on the uncorrelated excess map. 
This resulted in a best-fit position  $\alpha_{\mathrm{J2000}}=15h54m36s$,  $\delta_{\mathrm{J2000}}=-55^{\circ}05'05''$
(represented in Fig.~\ref{fig:rgb} in the multi-wavelength context)
with a statistical error on each coordinate of $\sim$1 arcmin
and an intrinsic Gaussian width (deconvoluted from the PSF) of (1.8 $\pm$ 0.6) arcmin.

Interestingly, the \g-ray centroid lies along the symmetry axis along the X-ray cometary structure, 
in an intermediate position between the pulsar candidate and the radio centroid. 


\subsection{Spectral results}

\begin{figure}
   \centering

   \includegraphics[bb= 56 350 560 717, clip,width=8.cm]{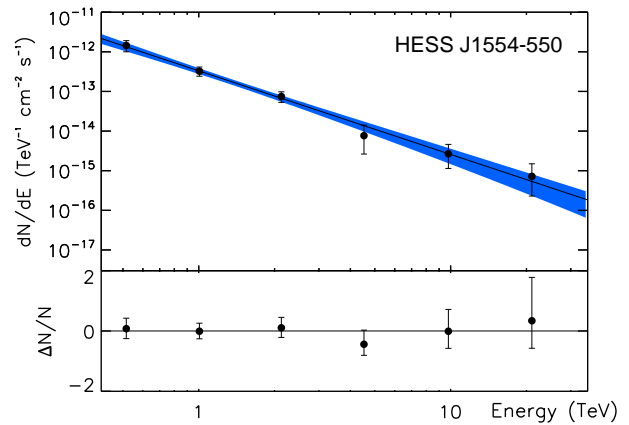}   
 
      \vspace{-0.2cm}

   \caption{Differential energy spectrum of the \g-ray source HESS J1554-550. The spectral data points, binned 
   in order to reach a significance level of 2$\sigma$, are well described by a power-law model (equivalent $\chi^{2}$/dof = 39.6/32) 
   whose best-fit is represented by a black line and residuals are shown in the lower panel.
    The blue band corresponds to the one sigma error of the spectral fit. }
   \label{fig:spectrum}
\end{figure}

The energy spectrum of the source was reconstructed with the
forward-folding maximum likelihood fit method \cite{piron}
 from a circular region of radius 0.1$^{\circ}$ chosen to fully encompass the 
emission from the object and centered on the \g-ray best-fit position previously derived.
 After background subtraction, using the \textit{multiple reflected-regions technique} \cite{berge07}, the  resulting
number of excess events is 109.
The spectrum, presented in Fig.~\ref{fig:spectrum}, can be well described by a power-law model
 (${\rm d}N/{\rm d}E = N_{\rm 0} (E/E_{\rm 0})^{-\Gamma} $) with a photon index $\Gamma = 2.1 \pm 0.2_{\rm stat} \pm 0.2_{\rm syst}$
and a normalization 
$N_{0} = (1.7  \pm 0.3_{\rm stat} \pm 0.4_{\rm syst} ) \times$ 10$^{-13}$ cm$^{-2}$s$^{-1}$ TeV$^{-1}$
given at the decorrelation energy $E_{\rm 0} = 1.39 $ TeV.
The integrated energy flux in the 1-10 TeV energy band is 
$ 1.12  \times10^{-12}$ ergs cm$^{-2}$ s$^{-1}$
and represents $\sim$1.5\% of the Crab nebula flux in the same energy band.

The \g-ray index derived here is almost identical to that measured in X-rays with \xmm~for the whole PWN
  (2.11$\pm$ 0.03 see Table 1 in \cite{temim09}), pointing towards the same population of electrons as origin of the synchrotron and the inverse Compton emissions seen in X-rays and \g-rays, respectively.

\section{Conclusions}


%

Observations with the \h~telescope array have led to the discovery of  hard TeV gamma-ray emission inside the composite SNR \snr. 
The HESS source is in spatial coincidence with the radio and X-ray PWN, and its centroid lies between the two. Although a bow-shock scenario 
(due to a supersonic pulsar velocity)  can not be excluded, the morphology of the overall system is reminiscent of that seen in other offset-type PWNe, such as HESS J1825-139 \cite{ah06b}, where the wind nebula has been crushed by an asymmetrical reverse shock.  
In contrast to many such PWNe, the composite SNR \snr~represents an excellent object to study the interaction between the PWN 
and its host and probe the system's evolutionary stage, as the SNR's shell is clearly identified and 
the characteristics of the PWN can now be studied in a broad multi-wavelength picture.



\begin{center}
\begin{small}ACKNOWLEDGMENTS          
\end{small}         
\end{center}

The support of the Namibian authorities and of the University of Namibia in facilitating the construction
and operation of H.E.S.S. is gratefully acknowledged, as is the support by the German Ministry for Education and Research (BMBF), the Max Planck Society, the French Ministry for Research, the CNRS-IN2P3 and the Astroparticle Interdisciplinary Programme of the CNRS, the U.K. Science and Technology Facilities Council (STFC), the IPNP of the Charles University, the Polish Ministry of Science and Higher Education, the South African Department of Science and Technology and National Research Foundation, and by the University of Namibia. We appreciate the excellent work of the technical support staff in Berlin, Durham, Hamburg, Heidelberg, Palaiseau, Paris, Saclay, and in Namibia in the
construction and operation of the equipment.

\clearpage

\end{document}